\documentclass[prl,aps,twocolumn,showpacs]{revtex4}
\usepackage{amssymb}
\usepackage{amsmath}
\usepackage{amsfonts}
\usepackage[dvips]{graphicx}

\begin{document}

\title{Photon-graviton pair conversion}
\author{Andreas K\"{a}llberg}
\author{Gert Brodin}
\email[E-mail: ]{gert.brodin@physics.umu.se}
\author{Mattias Marklund}
\email[E-mail: ]{mattias.marklund@physics.umu.se}
\affiliation{Department of Physics, Ume\aa\ University, SE-901 87 Ume\aa, Sweden}
\date{\today}

\begin{abstract}
We consider the conversion of gravitons and photons as a four-wave mixing process. A nonlinear coupled systems of equations involving two gravitons and two photons is obtained, and the energy exchange between the different degrees of freedom is found. The scattering amplitudes are obtained, from which a crossection for incoherent processes can be found. An analytical example is given, and applications to the early Universe are discussed.
\end{abstract}


\maketitle

The interaction between electromagnetic waves (EMWs) and gravitational waves (GWs) have been considered by many authors, using approximate methods \cite{Brodin00,Marklund00,Mendonca02,Servin03b,Vlahos04}, or by deriving exact solutions \cite{Griffiths91}. The later solutions are usually found on a flat vacuum background, in which case the simplest nonlinear interaction processes involving three waves are excluded due to energy-momentum conservation. In the present letter we will consider four-wave interaction on a Minkowski background, the lowest order process allowing for energy exchange between photons and gravitons. Using the classical Einstein-Maxwell system, we adopt a calculation scheme inspired by quantum mechanics \cite{DeWitt67I,DeWitt67II,DeWitt67III,Chen97,Ravndal02,Bastianelli05}. Assuming that two GWs interact with two EMWs, the only combination leading to energy exchange, the coupled system of equations is solved perturbatively, taking into account terms up to third order in the amplitudes. The four wave interaction coefficients (giving the quantum mechanical transition rates) are calculated in the center of mass system for arbitrary polarizations and scattering angle. Thus,  the local energy transfer rate between GWs and EMWs is found, i.e.\ the probability of photoproduction by two gravitons or that of the inverse process.

Although the time-scale for coherent GW-EMW pair conversion is shorter than the incoherent one, the large energy densities in the early universe makes it interesting to consider incoherent phenomena involving multiple conversion processes. Using the analogy between the interaction coefficients and the scattering matrix elements, the differential cross-section for graviton-photon pair conversion is calculated. In particular, it is found that a thermalization between photons and gravitons due to pair conversion requires a temperature $T\rightarrow T_p$, where $T_p$ is the Planck temperature.

In this letter, we use metric signature $(-+++)$, tetrad indices $a,b,\ldots=0,1,2,3$, $\alpha,\beta,\ldots=1,2,3$, and coordinate indices $\mu,\nu,\ldots=0,1,2,3$. The velocity of light in vacuum is $c=1$.

We use the formalism developed in Refs. \cite{Ellis99,Brodin00,Marklund00,Servin03b}, introducing an observer four-velocity, $V^{a}$, and decomposing the electromagnetic field into an electric and a magnetic part. Introducing an orthonormal frame $\left\{ \mathbf{e}_{a}=e_{a}^{\mu}\partial_{\mu}\right\}$, where $\mathbf{e}_{0}=\mathbf{V}=V^{a}\mathbf{e}_{a}$, (i.e. $V^{a}=\delta^{a}_{0}$), the Maxwell field equations can be written in terms of the electric and magnetic fields with the inclusion of effective charges and currents originating from the gravitational field (see e.g.\ \cite{Brodin00, Marklund00, Servin03b}).

From the field equations one may derive the following inhomogeneous wave equations
\begin{eqnarray}
\tilde\Box E^{\alpha}&=&-{\bf e}_0j^{\alpha}_{E}-\varepsilon^{\alpha\beta\gamma}{\bf e}_{\beta}j_{B\gamma}-\delta^{\alpha\gamma}{\bf e}_{\gamma}\rho_{E}\nonumber\\
& & -\varepsilon^{\alpha\beta\gamma}C^{a}_{\beta 0}{\bf e}_{a}B_{\gamma}-\delta^{\alpha\gamma}C^{a}_{\beta\gamma}{\bf e}_{a}E^{\beta}\label{ewav}\\
\tilde\Box B^{\alpha}&=&-{\bf e}_0j^{\alpha}_{B}+\varepsilon^{\alpha\beta\gamma}{\bf e}_{\beta}j_{E\gamma}-\delta^{\alpha\gamma}{\bf e}_{\gamma}\rho_{B}\nonumber\\
& & +\varepsilon^{\alpha\beta\gamma}C^{a}_{\beta 0}{\bf e}_{a}E_{\gamma}-\delta^{\alpha\gamma}C^{a}_{\beta\gamma}{\bf e}_{a}B^{\beta}\label{bwav}
\end{eqnarray}
where $\tilde\Box\equiv{\bf e}_{0}{\bf e}_{0}-\nabla\cdot\nabla$ and where the three-vector notation $\mathbf{E}\equiv (E^{\alpha })=(E^{1},E^{2},E^{3})$ etc. and $\nabla \equiv (\mathbf{e}_{1},\mathbf{e}_{2},\mathbf{e}_{3})$ from \cite{Brodin00, Marklund00, Servin03b} is used. The $C^{a}_{bc}$ are the commutation functions, i.e.\ $[{\bf e}_a,{\bf e}_b]
=C^{c}_{ab}{\bf e}_{c}$).

We consider plane waves with weakly space and time dependent amplitudes, i.e.\ $E=E(x^{\mu})\exp{[{\rm i}k_{\mu}x^{\mu}]}+c.c.$ and $|\partial_{\mu}E(x^{\nu})| \ll |k_{\mu}| |E(x^{\nu})|$, where $c.c.$ denotes the complex conjugate. In a flat vacuum background, energy-momentum conservation excludes three wave processes leading to energy exchange between photons and gravitons. Furthermore, the only four-wave combination converting energy between the gravitational and electromagnetic degrees of freedom involves two EMWs and two GWs. We denote the gravitational and electromagnetic waves by $h_{A}, h_{B}$ and $E_{A}, E_{B}$ respectively. The matching conditions for resonant four wave interaction are \cite{Weiland77} $k^{\mu}_{E_{A}}+k^{\mu}_{E_{B}}=k^{\mu}_{h_{A}}+k^{\mu}_{h_{B}}$, corresponding to energy-momentum conservation. In the center of mass system, all frequencies are equal and denoted by $\omega$. Thus, in this system we have ${\bf k}_{h_{B}}=-{\bf k}_{h_{A}}$, ${\bf k}_{E_{B}}=-{\bf k}_{E_{A}}$, fulfilling the matching conditions, and corresponding to pairwise counterpropagating waves \footnote{Apart from processes leading to GW-EM energy conversion, we can also have pure scattering processes where the EMW and GW energy are conserved separately, i.e. one photon and one graviton scatters into another photon-graviton pair. This process is not considered here.}. We consider terms up to cubic order in the amplitudes, and note that the corresponding effects can be grouped into four categories: 

\textit{i}) \textit{Nonresonant terms}. Most nonlinear combinations of wave amplitudes are nonresonant, and their effects vanishes after averaging over several wavelengths and time periods.

\textit{ii}) \textit{Frequency shift terms}. Various combination lead to (small) frequency shifts of the involved waves, described by interactions of the form $\Box E_{A}=Ch_{A}h_{A}^{\star}E_{A}+\dots$. However, these shifts are typically not significant, and the corresponding influence on the conversion process will be neglected.

\textit{iii}) \textit{Background curvature effects}. The energy momentum pseudo-tensor of each wave generates a slowly varying background curvature. The path of each wave will be described by (cubic) products of (second order) background curvature terms and linear terms. This curvature effect on photon-graviton conversion will be small provided the size of all the pulses are limited in all dimensions by $L$, fulfilling $L\ll\kappa c W /\omega$, where $W$ is the energy density of the pulses, and $\kappa$ is Einstein's gravitational constant.

\textit{iv}) \textit{Resonant energy conversion terms}. Cubic terms involving all wave amplitudes in accordance to the matching condition correspond to pair production of photons by two gravitons, or the converse process. 

We will only include the energy conversion terms from point \textit{iv} above, and the final interaction equations will then be of the form \cite{Weiland77}
\begin{equation}
\Box \Phi_{1}=M\Phi_{3}\Phi_{4}\Phi_{2}^{\star}\label{evoleqEa}
\end{equation}
and permutations thereof allowed by the wave matching.   
Here $\Phi_{1,2}$ and $\Phi_{3,4}$ represent mutually counterpropagating wave fields, (i.e.\  $E_{A,B}$ and $h_{A,B}$ respectively), $\Box\equiv\partial^{2}_{t}-\partial^{2}_{x}-\partial^{2}_{y}-\partial^{2}_{z}$, and the coefficients $M$ depend on the scattering angle, the polarization of the waves and on the frequency, $\omega$.

We introduce perturbations of the Minkowski metric, $g_{\mu\nu}=\eta_{\mu\nu}+h_{\mu\nu}$. The perturbed energy momentum tensor $T_{ab}$ should then satisfy Einstein's equations $G_{ab}=\kappa T_{ab}$, where $G_{ab}$ is the perturbation to the Einstein tensor due to the metric perturbation.

In order to determine the nonlinear gravitational response to the gravitational wave perturbations, we first consider the vacuum case. The perturbed metric is
\begin{equation}\label{vacmetric}
g_{\mu\nu}=\eta_{\mu\nu}+h_{\mu\nu}^{(1)}+h_{\mu\nu}^{(2)}
\end{equation}
where the $h_{\mu\nu}^{(1)}$ are wave perturbations to first order in amplitude and the $h_{\mu\nu}^{(2)}$ are the second order response to these perturbations. For the linear wave perturbations we choose the transverse and traceless gauge \cite{Landau75}, such that the only nonzero components are $h_{11}^{(1)}=-h_{22}^{(1)}\equiv h_{+}(t,z)$ and $h_{12}^{(1)}=h_{21}^{(1)}\equiv h_{\times}(t,z)$. 
To second order, the vacuum Einstein equations takes the form
$
R_{ab}^{(1)}+R_{ab}^{(2)}=0
$, 
where $R_{ab}^{(1)}$ is linear in $h_{+,\times}$ and $R_{ab}^{(2)}$ is quadratic in $h_{+,\times}$ and linear in $h_{\mu\nu}^{(2)}$. From the second order vacuum Einstein equations, we see that we only need the following components of $h_{\mu\nu}^{(2)}$, corresponding to a choice of gauge: $h_{00}^{(2)}$, $h_{03}^{(2)}$, $h_{11}^{(2)}$, $h_{22}^{(2)}$ and $h_{33}^{(2)}$. For the metric (\ref{vacmetric}) we choose the tetrad basis ${\bf e}_{a}^{(vac)}=e^{\mu(vac)}_{a}\partial_{{\mu}}=(e^{\mu(1)}_{a}+e^{\mu (2)}_{a})\partial_{{\mu}}$ where
\begin{eqnarray}
&&\!\!\!\!\!\!\!\!\!\!\!\!\!\!\!\!\! {\bf e}_0^{(vac)} = (1 + \tfrac{1}{2}{h_{00}^{(2)}})\partial_{t} - \tfrac{1}{2}{h_{03}^{(2)}}\partial_{z}\label{e0first}\\
&&\!\!\!\!\!\!\!\!\!\!\!\!\!\!\!\!\! {\bf e}_1^{(vac)} = (1 - \tfrac{1}{2}{h_{+}} - \tfrac{1}{2}{h_{11}^{(2)}} + \tfrac{1}{4}{h_{+}^{2}} + \tfrac{1}{4}{h_{\times}^{2}})\partial_{x} - \tfrac{1}{2}{h_{\times}}\partial_{y}\\
&&\!\!\!\!\!\!\!\!\!\!\!\!\!\!\!\!\! {\bf e}_2^{(vac)} = -\tfrac{1}{2}{h_{\times}}\partial_{x} + (1 + \tfrac{1}{2}{h_{+}} - \tfrac{1}{2}{h_{22}^{(2)}} + \tfrac{1}{4}{h_{+}^{2}} + \tfrac{1}{4}{h_{\times}^{2}})\partial_{y}\\
&&\!\!\!\!\!\!\!\!\!\!\!\!\!\!\!\!\! {\bf e}_3^{(vac)} = \tfrac{1}{2}{h_{03}^{(2)}}\partial_{t} + (1 - \tfrac{1}{2}{h_{33}^{(2)}})\partial_{z} . \label{e3first}
\end{eqnarray}
In this basis the only nonzero components of $R_{ab}^{(1)}$ are $R_{11}^{(1)}=-R_{22}^{(1)}=1/2(\partial_{t}^{2}-\partial_{z}^{2})h_{+}$ and $R_{12}^{(1)}=1/2(\partial_{t}^{2}-\partial_{z}^{2})h_{\times}$. The nonzero second order components are $R_{00}^{(2)}, R_{03}^{(2)}, R_{11}^{(2)}, R_{22}^{(2)}$ and $R_{33}^{(2)}$ where  $R_{ab}^{(2)}=R_{ab}^{(2)}(\partial^{2}h_{ab}^{(2)},\partial^{2}h_{+}^{2},\partial^{2}h_{\times}^{2})$. 
The second order nonlinear response terms $h_{\mu\nu}^{(2)}$ are related to the linear wave perturbations by $R_{ab}^{(2)}=0$, which gives $h_{11}^{(2)}=h_{22}^{(2)}$. Furthermore, adding $R_{00}^{(2)}$ to $R_{33}^{(2)}$, using the approximation $\partial_{t}\approx\pm\partial_{z}$ on $h_{+,\times}$, we get
\begin{equation}\label{h11eq}
\partial_{t}^{2}h_{11}^{(2)}=h_{+}\partial_{t}^{2}h_{+}+h_{\times}\partial_{t}^{2}h_{\times} .
\end{equation}
For two (counterpropagating) gravitational waves $h_{A}$ and $h_{B}$, we have  $h_{+}=h_{A+}+h_{B+}$ and $h_{\times}=h_{A\times}+h_{B\times}$, where the $h_{A}$ and $h_{B}$ are given by $h_{A}=h_{A}(t,z)\exp{[{\rm i}(k_{h}z-\omega t)]}+c.c. $ and $ h_{B}=h_{B}(t,z)\exp{[{\rm i}(-k_{h}z-\omega t)]}+c.c.$. 
From the form of the evolution equations (\ref{evoleqEa}), we see that the only terms contributing to the four wave interaction are terms proportional to $h_A h_B$ (and $h_A^{\star} h_B^{\star}$), i.e.\ terms $\propto \exp{[-2i\omega t]}$ (and $\exp{[2i\omega t]}$). By examining $R_{ab}^{(2)}$ we see that the only components containing terms with this purely temporal dependence are $h_{11}^{(2)}$, $h_{22}^{(2)}$ and $h_{33}^{(2)}$. 
Inserting the ansatz for $h_A$ and $h_B$ into Eq. (\ref{h11eq}), we get $h_{11}^{(2)}=(h_{+}^{2}+h_{\times}^{2})/4$, and the equation $R_{00}^{(2)}=0$ then gives $h_{33}^{(2)}=-h_{11}^{(2)}$. These second order quantities will enter the evolution equations (\ref{ewav}) and (\ref{bwav}), and contribute to the interaction coefficients in (\ref{evoleqEa}).

The directly resonant terms described above are not the only contribution to the coupling coefficients. Let ${\bf E}={\bf E}_{A}+{\bf E}_{B}=(E_{A1}+E_{B1}){\bf e}_{1}+(E_{A2}+E_{B2}){\bf e}_{2}+(E_{A3}+E_{B3}){\bf e}_{3}$ and similarly for ${\bf B}$. Furthermore, let $E_{A1}=E_{A1}(t,x,z)\exp{[{\rm i}(k_{Ex}x+k_{Ez}z-\omega t)]}+c.c. $, $E_{B1}=E_{B1}(t,x,z)\exp{[{\rm i}(-k_{Ex}x-k_{Ez}z-\omega t)]}+c.c.$ etc. Inserting this into the nonlinear wave equations for the electromagnetic fields, the largest terms on the right hand of Eqs.\ (\ref{ewav}) and (\ref{bwav}) are terms of the form $Eh, Bh$ etc., which induces oscillations at frequencies and wavenumbers that are nonresonant (nr), i.e.\ $\omega_{nr}\neq k_{nr}$. These nr fields may combine with fields of appropriate frequency and wavenumber in (\ref{ewav}) and the corresponding evolution equations for $h_A$, $h_B$, making the resulting terms resonant with the original wave perturbation, thus contributing to the coupling coefficients in (\ref{evoleqEa}).

In order to take these contributions into account we need to modify our ansatz for the electromagnetic field components, to include the nr fields. For this purpose, we use the lowest order approximation $\partial_{t}{\bf B}=-\nabla\times{\bf E}$, where $\nabla=(\partial_{x},\partial_{y},\partial_{z})$, in all second order terms in Eq.\  (\ref{ewav}). We thus make the ansatz $E_{1}^{(tot)}=E_{A1}+E_{B1}+E_{1}^{(nr)}$ etc.\ for the total electromagnetic field. 
Inserting this into Eqs.\ (\ref{ewav})--(\ref{bwav}), including terms that will contribute to the four wave coupling, we see that the terms linear in $h_{A,B}$ takes the form
$
\Box E^{(nr)}=c_1E_A^{\star}h_A+c_2E_A^{\star}h_B+c_3E_B^{\star}h_A+c_4E_B^{\star}h_B+c.c., \, 
$
 where the $c_{i}$ are coefficients determined by the field geometry. Thus, we find that the nr fields will contain oscillating factors $\exp{[{\rm i}({\bf k}_{E}-{\bf k}_{h})]}+\exp{[{\rm i}({\bf k}_{E}+{\bf k}_{h})]}+c.c.$. Applying this together with the previous expressions for the original perturbations, noting that for the given field geometry we may use $k_{Ez}=\omega\cos\theta$, $k_{Ex}=\omega\sin\theta$ and $k_h=\omega$, one obtains explicit expressions for the nr electric and magnetic field components. These expressions are greatly simplified when expressing the nr field components in terms of the linear polarization states for the original waves. We thus define the two independent linear electromagnetic polarization states according to: $E_{A1}=-\cos\theta E_{A+}$, $E_{A2}=E_{A\times}$, $E_{A3}=\sin\theta E_{A+}$, and $E_{B1}=\cos\theta E_{B+}$, $E_{B2}=E_{B\times}$, $E_{B3}=-\sin\theta E_{B+}$. The parts of the nr field components contributing to the four wave interaction take the form
\begin{eqnarray}
E_{1}^{(nr)}\!\!\! &=& \!\!\!\frac{1}{2}\cos\theta\left(-\cos\theta E_{A+}h_{A+}^{\star} + \cos\theta E_{B+}^{\star}h_{B+} \right.\nonumber\\
&+& \!\!\!\left. E_{A\times}h_{A\times}^{\star} + E_{B\times}^{\star}h_{B\times}\right)e^{{\rm i}({\bf k}_E-{\bf k}_h)\cdot {\bf r}}+c.c.
\end{eqnarray}
plus an analogous term with spatial dependence $\exp{[{\rm i}({\bf k}_{E}+{\bf k}_{h})]}$. 

Terms proportional to $E_{A}E_{B}$ and $E_{A}^{\star}E_{B}^{\star}$ oscillates as $\exp{(-2{\rm i}\omega t)}$ and $\exp{(2{\rm i}\omega t)}$, respectively. These terms induces nonresonant GW fields $h^{(nr)}$, with the same temporal variation, through Einstein's equations. Nonlinearities in the Einstein tensor of the form $h^{(nr)}h_{A,B}^{\star}$ are then resonant with the original perturbations, and will contribute to the coefficients in the evolution equations for the GW amplitudes.

Calculating the part of the energy momentum tensor varying as $\exp{(\pm 2{\rm i}\omega t)}$,  
we make the ansatz $E_{A1}=E_{A1}(t,x,z)\exp{[{\rm i}(k_{Ex}x+k_{Ez}z-\omega t)]}+c.c.$ etc. We have not included the nonresonant fields calculated above in the ansatz, since these fields do not appear to second order. 
The nonzero components of the energy momentum tensor varying as $\exp{(\pm 2{\rm i}\omega t)}$ are the purely spatial components of $T_{ab}$.

Since the energy-momentum tensor induces gravitational fields with the same temporal variation, $\exp{(\pm 2{\rm i}\omega t)}$, we assume a general metric perturbation with this time dependence. We can separate out terms in Einstein's equations describing the GWs and the nonlinear response to these, and terms describing the metric response to the energy momentum tensor of the EMWs, and therefore we can calculate the nr gravitational fields separately. We use the gauge $h^{(nr)}_{0\nu}=0$, 
and we choose a general tetrad (i.e.\ a basis where all $h_{\alpha\beta}^{(nr)}$ are included),  
such that the extra terms corresponding to the nr fields basis may be added to the tetrad (\ref{e0first})--(\ref{e3first}). We only need to include terms up to ${\cal O} (h^{(nr)}_{\mu\nu})$ in the Einstein tensor, since higher order terms in $h^{(nr)}_{\mu\nu}$ will be of at least fourth order in the electromagnetic amplitude. The nonzero components of the Einstein tensor to this order are the purely spatial components of $G_{ab}^{(nr)}$. By letting $h^{(nr)}_{\alpha\beta} \rightarrow h^{(nr)}_{\alpha\beta}\exp{[-2{\rm i}\omega t]}+c.c.$ one obtains explicit expressions for the nonresonant gravitational fields from Einstein's equations.

We now have the ingredients needed to calculate the evolution equations for the amplitudes of the waves and explicitly give the coefficients in (\ref{evoleqEa}). We expand our metric ansatz (\ref{vacmetric}), in order to include the nonresonant GW fields, as $g_{\mu\nu}=\eta_{\mu\nu}+h_{\mu\nu}^{TT}+h_{\mu\nu}^{(2)}+h_{\mu\nu}^{(nr)}$, and include the driven nonresonant part in our ansatz for the electromagnetic fields. The final tetrad basis is given by ${\bf e}_{a}=(e_{a}^{\mu (vac)}+e_{a}^{\mu (nr)})\partial_{{\mu}}$, where
\begin{equation}
e^{\mu (nr)}_{a} = \left( \begin{array}{cccc}
0 & 0 & 0 & 0\\
0 & -{\tfrac{1}{2}{h_{11}^{(nr)}}} & -{\tfrac{1}{2}{h_{12}^{(nr)}}} & -{\tfrac{1}{2}{h_{13}^{(nr)}}}\\
0 & -{\tfrac{1}{2}{h_{12}^{(nr)}}} & -{\tfrac{1}{2}{h_{22}^{(nr)}}} & -{\tfrac{1}{2}{h_{23}^{(nr)}}}\\
0 & -{\tfrac{1}{2}{h_{13}^{(nr)}}} & -{\tfrac{1}{2}{h_{23}^{(nr)}}} & -{\tfrac{1}{2}{h_{33}^{(nr)}}}
\end{array}\right) ,
\end{equation}
and where we put $h_{00}^{(2)}=h_{03}^{(2)}=0$ in $e_{a}^{\mu(vac)}$. Using this basis, and inserting explicit expressions for all fields into Eq.\ (\ref{ewav}), one obtains equations, involving several time scales, for the field components.   
Taking an average, over several wavelengths and periods, of the equations, terms resonant with the original perturbation are obtained 
 \cite{Weiland77}. Expressing the equations using the linear polarization states, we finally obtain 
\begin{eqnarray}
\Box E_{A\scriptscriptstyle\overset{+}{\times}} &=& M_{+}H_{I}E_{B\scriptscriptstyle\overset{+}{\times}}^{\star} \pm M_{\times}H_{II}E_{B\scriptscriptstyle\overset{\times}{+}}^{\star}\label{Eaeq1} ,\\
\Box E_{B\scriptscriptstyle\overset{+}{\times}} &=& M_{+}H_{I}E_{A\scriptscriptstyle\overset{+}{\times}}^{\star}\mp M_{\times}H_{II}E_{A\scriptscriptstyle\overset{\times}{+}}^{\star}\label{Ebeq1} ,
\end{eqnarray}
from Eq. (\ref{ewav}), where $H_{I} \equiv h_{A+}h_{B+}-h_{A\times}h_{B\times}$, $H_{II} \equiv h_{A+}h_{B\times}+h_{A\times}h_{B+}$, $M_{+}\equiv \omega^2(1+\alpha^2)/2$, $M_{\times}\equiv \omega^2\alpha$ and $\alpha \equiv \cos\theta$.

The evolution equations for the gravitational wave amplitudes are $G_{11}-G_{22} = \kappa( T_{11}-T_{22})$ for $h_{+}$ and $G_{12}+G_{21} = \kappa(T_{12}+T_{21})$ for $h_{\times}$. Inserting explicit expressions for all fields, 
we obtain the following equations
\begin{eqnarray}
\Box h_{A\scriptscriptstyle\overset{+}{\times}} &=& \pm m_{+}E_{I}h_{B\scriptscriptstyle\overset{+}{\times}}^{\star}+m_{\times}E_{II}h_{B\scriptscriptstyle\overset{\times}{+}}^{\star}\label{haeq1}\\
\Box h_{B\scriptscriptstyle\overset{+}{\times}} &=& \pm m_{+}E_{I}h_{A\scriptscriptstyle\overset{+}{\times}}^{\star}+m_{\times}E_{II}h_{A\scriptscriptstyle\overset{\times}{+}}^{\star}\label{hbeq1}
\end{eqnarray}
where $E_{I} \equiv  E_{A+}E_{B+}+E_{A\times}E_{B\times}$, $E_{II} \equiv  E_{A+}E_{B\times}-E_{A\times}E_{B+}$, $m_{+}\equiv\kappa(1+\alpha^2)$ and $m_{\times}\equiv 2\kappa\alpha$. The coupled system of equations (\ref{Eaeq1})--(\ref{hbeq1}) describes the evolution of the slowly varying wave amplitudes.

Introducing the energy densities of the waves ${\cal E}_{GW}\approx\omega^2/2\kappa\left( \vert h_{+}\vert^2+\vert h_{\times}\vert^2\right)$ and ${\cal E}_{EM} = \vert E_{+}\vert^2+\vert E_{\times}\vert^2$ for the gravitational and electromagnetic wave respectively, we note that the total wave energy, ${\cal E}_{tot}$, is conserved in the interaction described by Eqs. (\ref{Eaeq1})--(\ref{hbeq1}) .

In order to illustrate the physics of (\ref{Eaeq1})--(\ref{hbeq1}), we assume long pulses, i.e.\ $\Box\rightarrow -2{\rm i}\omega\partial_t$.  For simplicity, we consider the case where all waves have +-polarization, and assume that the initial amplitudes and relative phases of the wave envelopes are fixed in such a way that $E_{A+} = E_{B+} \equiv E$ and $h_{A+} = h_{B+} \equiv h$ where $E=\hat E(t)\exp{[{\rm i}\varphi_{E}(t)]}$ and $h=\hat h(t)\exp{[{\rm i}\varphi_{h}(t)]}$. We insert this into Eqs.\ (\ref{Eaeq1}) and (\ref{haeq1}), and reexpress the equations as evolution equations for the energy densities ${\cal E}_{EM}$ and ${\cal E}_{GW}$, and the relative phases of the waves. Then, for the special choices of initial values $(\varphi_{h}-\varphi_{E})\vert_{t=0}=\pm\pi/4$, one obtains very simple analytical solutions for the ratio of electromagnetic and gravitational wave energy densities. The results in this case are ${\cal E}_{GW}/{\cal E}_{EM}=({\cal E}_{GW}/{\cal E}_{EM})(0)e^{-\tau}$ in the case when $(\varphi_{h}-\varphi_{E})(0)=-\pi/4$ and ${\cal E}_{EM}/{\cal E}_{GW}=({\cal E}_{EM}/{\cal E}_{GW})(0)e^{-\tau}$ for $(\varphi_{h}-\varphi_{E})(0)=\pi/4$, where $\tau\equiv(1+\alpha^2)\kappa{\cal E}_{tot}t/2\omega$.

Thus we see that the characteristic time-scale for graviton-photon 
conversion is rather slow. In particular, as far as the interaction 
between four long pulses is concerned, there is only time for large 
energy conversion if the waves also contains enough energy to 
significantly modify the background curvature (c.f.\ point \textit{iii}\ 
above). However, we note that the GW-EMW conversion takes place in a 
space-time that is locally flat, and a given pulse (EMW or GW) can 
interact with other pulses (or even individual quantas) consecutively, 
leading to significant conversion after multiple processes. Thus, we here 
emphasize that the coefficients in Eqs.\ (\ref{Eaeq1})--(\ref{hbeq1}) can be used also to describe incoherent interaction involving multiple processes. Writing the equations in terms of the vector potential, $A$, rather than the electric field and making a normalization $h\rightarrow h/\sqrt{2\kappa}$, we note that $M_{+}, m_{+}\rightarrow{\cal M}_{+}\equiv\kappa\omega^2(1+\alpha^2)$ and $M_{\times}, m_{\times}\rightarrow{\cal M}_{\times}\equiv 2\kappa\omega^2\alpha$ in (\ref{Eaeq1})--(\ref{hbeq1}). The coefficients ${\cal M}_{+}$ and ${\cal M}_{\times}$ now correspond to the scattering amplitude matrix elements, from which the cross-section can be calculated straightforwardly. Following Ref.\ \cite{Itzykson85}, we find that the differential cross-section for graviton-photon conversion averaged over all polarization states is
\begin{eqnarray}    
\frac{d\sigma}{d\Omega} = \frac{\vert{\cal M}_{+}\vert^2+\vert{\cal M}_{\times}\vert^2}{128\omega^2(2\pi)^2}
= \left[\frac{(1 + \alpha^2)^2}{8} + \frac{\alpha^2}{2}\right]L_P^4\omega^2
\label{crossect}
\end{eqnarray}
where $L_P$ is the Planck length.

The process of graviton--photon conversion may lead to thermalization between gravitons and photons in the early universe. For thermalization to occur before expansion slows down the conversion rate too much, the collision frequency must fulfill $\nu = \sigma n > H$, where $H$ is the Hubble parameter and $n$ is the number density. Considering a photon gas at temperature $T$ in the radiation dominated era, using $H \sim 1/t$, the above condition can be written
$
T/T_P \gtrsim \mathcal{N}^{1/6}_{\text{eff}}
$
where $T_P$ is the Planck temperature and $\mathcal{N}_{\text{eff}}$ is the effective number of degrees of freedom. Thus thermalization, in case this is still a meaningful concept when $T\sim T_P$, must be described within a quantum theory of gravity. The classical photon-graviton conversion process nevertheless puts some limits on the background levels of gravitational radiation, since it makes a relic GW spectrum much above the thermal level unlikely.

We note, however, that the time-scale for coherent interaction is shorter than that for incoherent single particle interactions by a factor of the order $k^2L_P^2$, and thus physical scenarios with significant GW-EMW conversion in 
astrophysics is still possible within a classical framework. Due to the large energy densities required, however, such processes are likely to take place in significantly curved backgrounds, for which a description lies outside the scope of this letter.

A comparison of the cross-section (\ref{crossect}) to that obtained by using Feynman diagrams would be of much interest, but as far as we know, the later result is yet to be deduced. We note, however, that an interaction lagrangian of the form ${\cal L}\propto h^{\mu\nu}T_{\mu\nu}$, as used in e.g. Ref.\ \cite{Ravndal02} in linear quantum gravity, will not be sufficient here, since the effect considered in this letter is quadratic in the gravitational wave amplitudes.

\acknowledgments{Andreas K\"allberg, who was supported by the Swedish National Graduate 
School of Space Technology, was tragically killed in an accident shortly after this work was completed.}

\bibliographystyle{apsrev}
\bibliography{cqg}

\end{document}